\documentclass[a4paper,12pt]{article}
\usepackage{psfig}
\usepackage{amssymb}
  
\newcommand{\gsim}{\raisebox{-0.5ex}{$\stackrel{>}{\sim}$}}
\newcommand{\lsim}{\raisebox{-0.5ex}{$\stackrel{<}{\sim}$}}   

\def\bc{\begin{center}}
\def\ec{\end{center}}

\def\be{\begin{equation}}
\def\ee{\end{equation}}
\def\beq{\begin{eqnarray}}
\def\eeq{\end{eqnarray}}
\def\bfig{\begin{figure}}
\def\efig{\end{figure}}
\def\bnum{\begin{enumerate}}
\def\enum{\end{enumerate}}
 
\begin{document}

\begin{flushright}
Journal-Ref: Astronomy Letters, 2005, v. 31, No 1, pp. 15-29
\end{flushright}

\vspace{1cm}
  
\bc
{\LARGE\bf Bending Instability of Stellar Disks:
The Stabilizing Effect of a Compact Bulge}\\
\vspace{0.7cm}
{\bf N. Ya. Sotnikova and S. A. Rodionov}\\
\vspace{0.7cm}
{\it Astronomical Institute, St. Petersburg State University,
Universitetskii pr. 28, 198904 Russia}\\
Received January 19, 2004
\ec

\vspace{1cm}
\bc \bf Abstract \ec

\noindent
The saturation conditions for bending modes in inhomogeneous thin stellar
disks that follow from an analysis of the dispersion relation are compared
with those derived from $N$-body simulations. In the central regions of
inhomogeneous disks, the reserve of disk strength against the growth of
bending instability is smaller than that for a homogeneous layer. The
spheroidal component (a dark halo, a bulge) is shown to have a stabilizing
effect. The latter turns out to depend not only on the total mass of the
spherical component, but also on the degree of mass concentration toward the
center. We conclude that the presence of a compact (not necessarily massive)
bulge in spiral galaxies may prove to be enough to suppress the bending
perturbations that increase the disk thickness. This conclusion is
corroborated by our $N$-body simulations in which we simulated the evolution
of almost equilibrium, but unstable finite-thickness disks in the presence
of spheroidal components. The final disk thickness at the same total mass of
the spherical component (dark halo $+$ bulge) has been found to be much
smaller than that in the simulations where a concentrated bulge is present.


\newpage
\section{INTRODUCTION}

A peculiarity of the disks in spiral galaxies is that these are rather thin
objects. Their thickness is several times smaller than the radial scale
length. How far stars can go from the principal galactic plane due to their
vertical random velocity component determines the disk thickness at fixed
star surface density. The larger the velocity dispersion, the thicker the
disk. The random velocities of young stars are known to be low, but the
stellar ensemble can subsequently heat up through various relaxation
processes; i.e., the random velocity dispersion can increase. Thus, the
stellar disk thickness depends on how effective the relaxation processes are
in galaxies, and it is ultimately determined by the factors that suppress or
trigger the various heating mechanisms. Three basic stellar disk heating
mechanisms are commonly discussed in the literature: the scattering of stars
by giant molecular clouds (Spitzer and Schwarzschild 1951, 1953), the
scattering by transient spiral density waves (among the first results of
numerical simulations are those obtained by Sellwood and Carlberg (1984)),
and the heating of the ensemble of stars that constitute the disks of spiral
galaxies as they interact with external sources, for example, with low-mass
satellites (see, e.g.,Walker {\it et al.} 1999; Velasquez and White 1999).

A second remarkable peculiarity of the stellar disks is that their structure
is unusually ``fragile''. This peculiarity was revealed by a linear analysis
of the collisionless Boltzmann equation and has been repeatedly illustrated
by $N$-body simulations. Numerous studies have shown that the initially
regular structure of the stellar disks can change radically due to the
growth of various instabilities, which give rise to large-scale structures
both in the disk plane (bars, spiral arms, rings) and in the vertical
direction (warps). The analytically and numerically obtained saturation
conditions for unstable modes impose the most severe restrictions on the
global structural and dynamical parameters of the stellar disks.

A local analysis suggests that the stars at a given distance $R$ from the
disk center must have a radial velocity dispersion $\sigma_R(R)$ larger than
some minimum critical value $\sigma^{\rm cr}_R(R)$ (Toomre 1964) for the
disk to be gravitationally stable in the region under consideration (at
least against the growth of axisymmetric perturbations). In addition, for
the disk to be stable against the growth of bending perturbations, the ratio
of the vertical and radial velocity dispersions $\sigma_z / \sigma_R$ must
also be larger than some value approximately equal to $0.2 - 0.37$ (Toomre
1966; Kulsrud {\it et al.} 1971; Polyachenko and Shukhman 1977; Araki 1985).
The latter quantity determines the minimum thickness of the galactic disk, and
its comparison  with the observed value allows the contribution of the
bending instability to the vertical disk heating to be estimated and
compared with the contribution of other relaxation mechanisms.

On the other hand, if we exclude the heating mechanisms mentioned above from
our analysis and take, as is commonly done, the condition for marginal disk
stability against the growth of bending modes by fixing 
$\sigma_z / \sigma_R$ at a level of the linear approximation, $0.29 - 0.37$,
then the velocity dispersion $\sigma_z$ at the same star surface density
will decrease with decreasing $\sigma_R$ (see, e.g., Zasov {\it et al.}
1991). In this case, the disk will have a smaller thickness.

The presence of a spheroidal component, for example, a dark halo is known to
produce a stabilizing effect and to decrease the minimum value of
$\sigma^{\rm cr}_R$ required for gravitational stability. Consequently, the
disks embedded in a massive halo, on average, must have low values of
$\sigma_z$ and be, on average, thinner. Based on similar reasoning, Zasov et
al. (1991, 2002) showed that the relative disk thickness $z_0/h$ ($z_0$ is
the half-thickness of the disk, and $h$ is the radial exponential scale
length) is proportional to $M_{\rm d} / M_{\rm t}$, where $M_{\rm d}$ and
$M_{\rm t}$ are, respectively, the mass of the disk and the total mass of
the galaxy within a fixed radius. These authors also concluded that a small
disk thickness suggests the existence of a massive dark halo in the galaxy.
Moreover, based on $N$-body simulations, Zasov {\it et al.} (1991) and 
Mikhailova {\it et al.} (2001) constructed a dependence that allows the
relative mass of the dark halo $M_{\rm h} / M_{\rm d}$ to be estimated from
$z_0/h$. However, as we show below, the relationship between the relative
disk thickness and the mass of the spheroidal component is more complex.

In this paper, we numerically analyze the saturation conditions for bending
instability in inhomogeneous three-dimensional stellar disks at nonlinear
stages in the presence of a spheroidal component of different nature (a
stellar bulge and a dark halo) and the constraints imposed on the final disk
thickness.

\section{PECULIARITIES OF THE GROWTH OF BENDING INSTABILITY IN INHOMOGENEOUS
THIN STELLAR DISKS}

\subsection{Global Modes}

To understand how the growth of bending instability in inhomogeneous disks
differs from that in homogeneous disks, let us first turn to the result of
Toomre (1966). Toomre was the first to derive the dispersion relation for
long-wavelength bending perturbations in an infinitely thin gravitating
layer with a nonzero velocity dispersion
\be
\omega^2 = 2 \pi G \Sigma |k| - \sigma_x^2 k^2 \, .
\label{dispersion1}
\ee
where $\Sigma$ is the star surface density of the layer, and $\sigma_x^2$ is
the velocity dispersion along a particular coordinate axis in the plane of
the layer. It follows from Eq.~(\ref{dispersion1}) that the perturbations
with a wavelength 
$\lambda = 2 \pi / k > \lambda_{\rm J} = \sigma_x^2 / G \Sigma$ are stable,
since $\omega^2 >0$ in this range. 

Relation~(\ref{dispersion1}) was derived for an infinitely thin disk and,
when applied to finite-thickness disks, is valid only for perturbations with
a wavelength longer than the vertical scale height of the system, i.e., for
$\lambda >> z_0$, where $z_0$ is the half-thickness of the layer. It can be
shown from general considerations that the longest-wavelength perturbations
in finite-thickness disks with a wavelength
$
\displaystyle \lambda < \lambda_2 \approx z_0 \frac{\sigma_x}{\sigma_z}
$
are stable, since they are smeared by thermal motions in the plane of the
layer (see, e.g., Polyachenko and Shukhman 1997). Having derived the
corresponding dispersion relation, Polyachenko and Shukhman (1977) were the
first to find the exact location of the stability boundary in the
short-wavelength range for a homogeneous flat finite-thickness layer. Araki
(1985) (see also Merritt and Sellwood 1994) obtained a similar result for a
homogeneous layer with a vertical density pro?le close to the observed one
in real galaxies\footnote{This profile corresponds to the model of an
isothermal layer (Spitzer 1942) and describes well the vertical density
variations observed in galaxies (van der Kruit and Searle 1981).} 
$\rho(R,z) = \rho(R,0) {\rm sech}^2(z/z_0)$. As regards the
intermediate-wavelength ($\lambda_2 < \lambda < \lambda_{\rm J}$)
perturbations, they are unstable, as follows from the results of the studies
mentioned above. 

As the disk thickness $z_0$ increases, the wavelength $\lambda_2$ increases
and tends to $\lambda_{\rm J}$. When $\lambda_2 = \lambda_{\rm J}$, the disk
is stabilized against bending perturbations of any wavelengths. The
following analytical estimate in the linear approximation obtained both from
qualitative considerations (Toomre 1966; Kulsrud {\it et al.} 1971) and from
an accurate analysis of the dispersion relation for a finite-thickness layer
(Polyachenko and Shukhman 1977; Araki 1985) is valid:
\be
\displaystyle
\left( \sigma_z / \sigma_x \right)_{\rm cr} \approx 0.29 - 0.37\, .
\label{crit1}
\ee
The instability is completely suppressed if 
$\sigma_z / \sigma_x > \left( \sigma_z / \sigma_x \right)_{\rm cr}$ and
grows if $\sigma_z / \sigma_x < \left( \sigma_z / \sigma_x \right)_{\rm cr}$. 

As regards the inhomogeneous models, the first thing that radically
distinguishes the growth of bending instability in inhomogeneous disks from
that in a homogeneous layer is the existence of global unstable bending
modes with a wavelength longer than the disk scale length. This conclusion
follows from an analysis of the equation that describes the evolution of
long-wavelength bending perturbations in an in infinitely thin disk with a
radially decreasing density (Polyachenko and Shukhman 1977; Merritt and
Sellwood 1994). For special disk models, it was shown that the region of
stable long-wavelength perturbations narrows significantly in this case
(see, e.g., Fig. 2 from Merritt and Sellwood (1994)). This is attributed to
the fact that the restoring force from the perturbation that grows in an
inhomogeneous disk (Merritt and Sellwood 1994) or in a radially bounded disk
(Polyachenko and Shukhman 1977) is always weaker than the corresponding
restoring force in a homogeneous infinite layer. This fact was demonstrated
more clearly by Sellwood (1996), who noted that the dispersion 
relation~(\ref{dispersion1}) 
could be used to analyze the bending instability in
inhomogeneous disks (at least qualitatively) if another term related to the
restoring force from the unperturbed disk is added to it:
\be
\omega^2 = \nu_{\rm d}^2 + 2 \pi G \Sigma |k| - \sigma_x^2 k^2 \, ,
\label{dispersion2}
\ee
where $\nu_{\rm d} = \sqrt{\partial^2 \Phi_{\rm d}(R,z) / \partial z^2}$ is
the vertical oscillation frequency of the stars, and $\Phi_{\rm d}(R,z)$ is
the potential of the disk.

For disks with a nonflat rotation curve, the additional term $\nu_{\rm d}^2$
can play a destabilizing role. As was noted by Sellwood (1996), for an
infinitely thin inhomogeneous disk,
\be
\nu_{\rm d}^2 =
\left.\frac{\partial^2 \Phi_{\rm d}(R,z)}{\partial z^2}\right|_{z=0+} =
- \left.\frac{1}{R} \frac{\partial}{\partial R}
\left(R \frac{\partial^2 \Phi_{\rm d}}{\partial R}\right)\right|_{z=0} =
- \frac{1}{R} \frac{d v_{\rm c,d}^2}{d R} \, ,
\label{rotcurve}
\ee
($v_{\rm c,d}$ is the circular rotational velocity of the disk), and
$\nu_{\rm d}^2 <0$ for $d v_{\rm c,d}^2 / d R > 0$.

Thus, an additional expulsive force from the unperturbed disk emerges in the
central regions where the rotation curve rises. The destabilizing effect of
the disk in the central regions gives rise to another region at small wave
numbers (long $\lambda$) where $\omega^2 < 0$. This region is responsible
for the growth of largescale bending instability and the emergence of global
modes. In this case, one might expect a larger disk thickness and a
larger value $\sigma_z / \sigma_x$ ($\sigma_z / \sigma_R$) than those
in homogeneous models to be required to suppress the instability.

Applying all of this reasoning to finite-thickness disks is not quite
obvious. On the one hand, it may be assumed that the restoring force from
the unperturbed disk $F_z = - \partial \Phi_{\rm d} / \partial z$  for 
$R \to 0$ behaves as $F_z \simeq - G M_{\rm d} z/|z|^3$ (here, $M_{\rm d}$
is the total mass of the disk) starting from some $z$, i.e., decreases (in
magnitude) with increasing $z$. Consequently, 
$\nu_{\rm d}^2 = \partial^2 \Phi_{\rm d} / \partial z^2 =
- \partial F_z / \partial z $ becomes negative (Sellwood 1996).

\begin{figure}
\centerline{\psfig{file=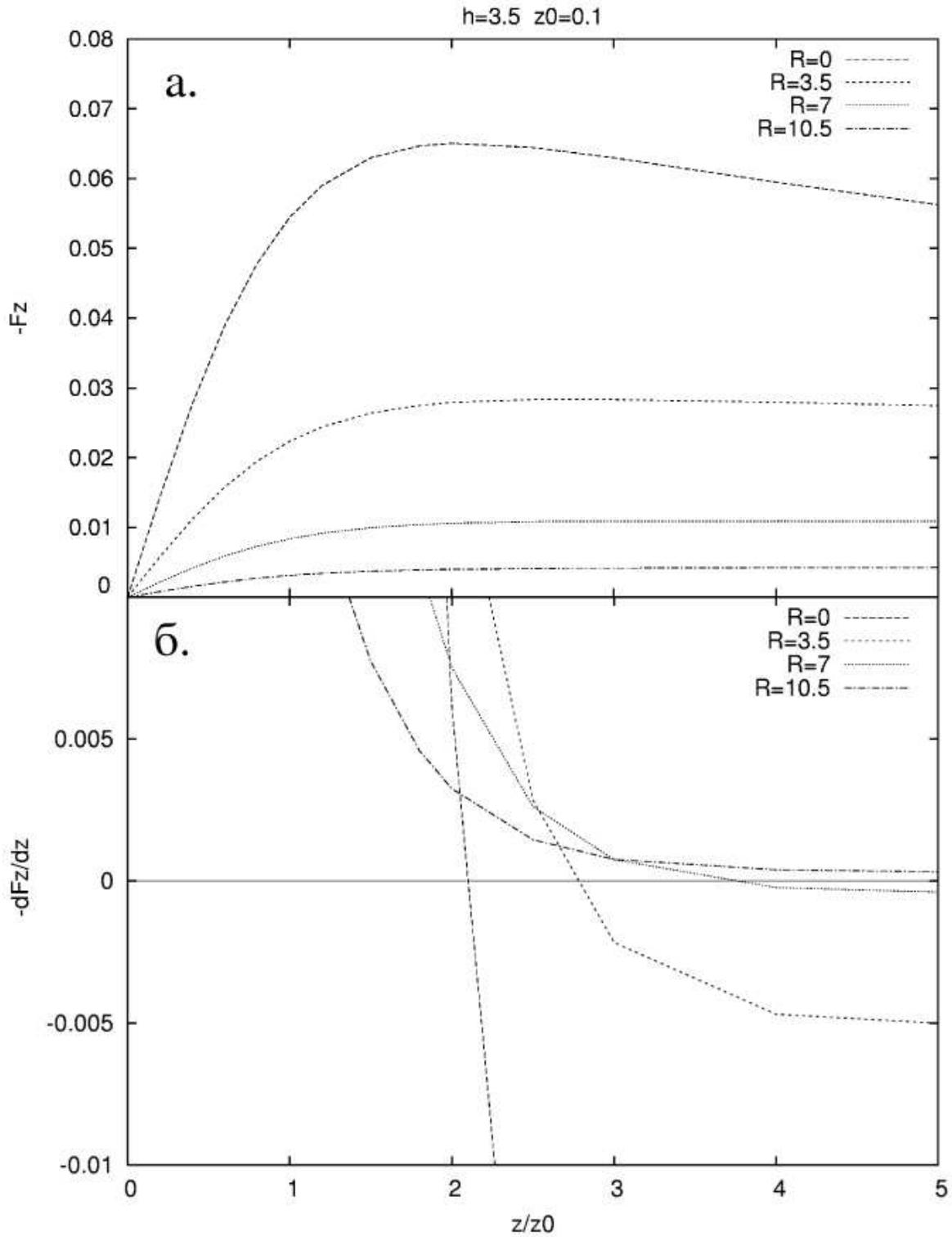,width=16cm}}
\caption[1]{(a) Magnitude of the vertical gravitational force
($ - F_z$) versus $z$ at various distances $R$ from the disk center;
(b) the square of the vertical oscillation frequency
$\nu_{\rm d}^2 = - \partial F_z / \partial z$ versus $z$ for various $R$. We
took $G = 1$ and the disk parameters $M_{\rm d} = 1$ (the total disk mass),
$h = 3.5$, and $z_0 = 0.1$.}
\label{force_z}
\end{figure}

On the other hand, we can exactly calculate the $z$ dependence of $\nu_{\rm
d}^2$ for a given R from the general formula
\be
\nu_{\rm d}^2({\bf r}) = - \partial F_z / \partial z =
- \frac{\partial}{\partial z} \int
\frac{G \rho_{\rm d}({\bf r^{\prime}})
(z^{\prime} - z)}{|{\bf r}^{\prime} - {\bf r}|^3} \,\,
d^3 {\bf r}^{\prime}  \, ,
\label{nu}
\ee
for the density profile that is commonly used to describe the disks of spiral
galaxies:
\be
\label{eq_star_disk_dens}
\rho_{\rm d}(R,z)=\frac{M_{\rm d}}{4 \pi h^2 z_0}
                     \cdot \exp\left(-\frac{R}{h}\right)
		                          \cdot {\rm sech}^2\left(
\frac{z}{z_0} \right)
\ee
where $h$ is the exponential scale length of the disk, $z_0$ is the vertical
scale height, and $M_{\rm d}$ is the total mass of the disk. The derived
dependence\footnote{We see from Eq.~(\ref{nu}) that triple integrals over
infinite intervals must generally be calculated to determine 
$\nu_{\rm d}^2$. This can be easily done by using adaptive algorithms for
calculating the integrals. In our integration,we used the gsl library
(information about the gsl (GNU Scientific Library) project can be found at
http://sourses/redhat/com/gsl). For models with known analytical
density-potential pairs, for example, the Miyamoto-Nagai disk (see Binney
and Tremaine (1987), p. 44), a comparison of the analytical $z$ dependence
of $\nu_{\rm d}^2$ at a given $R$ with that calculated numerically by triple
integration yielded a close match.} for various $R$ is shown 
in Fig.~\ref{force_z}b. Figure~\ref{force_z}a shows the $z$ dependence of the
vertical force $F_z$, which necessarily has an extremum at some $z$. In the
central regions, $\nu_{\rm d}^2$ is negative at $z \gsim 2z_0$ (on the
periphery, the passage to the negative region occurs at even larger 
$z / z_0$). This implies that all of the above reasoning for finitethickness
disks is directly applicable only to largeamplitude perturbations. It is
also clear that the bellshaped (or axisymmetric) mode with the azimuthal
number $m = 0$ must be most unstable in the central regions. If the
amplitude of the axisymmetric bend increases significantly during the growth
of bending instability, then it can raise the central stars above the plane
of the disk and bring them into a ``dangerous'' zone where 
$\nu_{\rm d}^2 < 0$; subsequently, the self-gravity of the disk itself will
contribute to the growth of a bend to the point of saturation.

\subsection{The Central Regions of Hot Infinitely Thin Disks}

The second peculiarity of the growth of bending instability in inhomogeneous
infinitely thin disks is as follows. The degree of disk instability against
bending perturbations depends on the degree of disk ``heating'' in the
plane. Exact and approximate analyses of the dispersion relation for an
inhomogeneous infinitely thin disk (Polyachenko and Shukhman 1977; Merritt
and Sellwood 1994) show that the modes with increasingly large azimuthal
numbers in the central regions become unstable for all wavelengths as the
fraction of the kinetic energy contained in the random motions in the disk
plane increases. The conclusion about the existence of a Toomre parameter  
$Q_{\rm T}$ (Toomre 1964), which characterizes the degree of heating of the
stellar disk in the plane, at which the disk cannot be stabilized against
bending perturbations of any wavelengths can also be drawn from 
Eq.~(\ref{dispersion2}), i.e., the equation written for long-wavelength
perturbations in a homogeneous layer, but ``rectifided'' by the term 
$\nu_{\rm d}^2$ to include the inhomogeneity effects. 

Indeed,  $\omega^2 <0$ for any $k$ if
$$
\nu_{\rm d}^2 + \frac{\left(\pi G \Sigma\right)^2}{\sigma^2_x} <0 \, .
$$
Given~(\ref{rotcurve}),
$$
\nu_{\rm d}^2 =
2 \Omega^2_{\rm d} - \kappa^2_{\rm d} \, ,
$$
where $\displaystyle \Omega_{\rm d} = \frac{v_{\rm c,d}}{R}$ 
is the angular velocity, and 
$\displaystyle \kappa^2_{\rm d} =
2 \frac{v_{\rm c,d}^2}{R^2}
\left(1 + \frac{R}{v_{\rm c,d}} \frac{d v_{\rm c,d}}{d R}\right)$ 
is the square of the epicyclic frequency. Then,
\be
\omega^2 <0 \mbox{\rm \,\,\,\,\,\, при \,\,\,\,\,\,}
2 \Omega^2_{\rm d} - \kappa^2_{\rm d} +
\frac{\left(\pi G \Sigma\right)^2}{\sigma^2_x} <0 \, .
\label{kat}
\ee
For the central regions of a rigidly rotating disk 
($\kappa^2_{\rm d} = 4 \Omega^2_{\rm d}$), we obtain from~(\ref{kat})
$$
\omega^2 <0 \mbox{\rm \,\,\,\,\,\, при \,\,\,\,\,\,}
\sigma_x > \sqrt{2} \, \frac{\pi G \Sigma}{\kappa_{\rm d}}
\approx \sqrt{2} \, \sigma_R^{\rm cr} \, ,
$$
or
\be
\omega^2 <0 \mbox{\rm \,\,\,\,\,\, при \,\,\,\,\,\,}
Q_{\rm T} > \sqrt{2} \mbox{\rm \vspace{1cm} for any wavelength,}
\label{kat2}
\ee
where $Q_{\rm T} = \sigma_x / \sigma_R^{\rm cr}$ is the Toomre
parameter\footnote{Actually, the Toomre parameter should have been defined as
$\sigma_R / \sigma_R^{\rm cr}$, but the difference does not matter in our
case.} (Toomre 1964).

Condition~(\ref{kat2}) is not an exact criterion; it is only an estimation
relation. However, it shows that the central regions of hot stellar disks
($Q_{\rm T} >> 1$) with a large reserve of strength against the growth of
instabilities in the plane of the disk (bars, spiral arms) cannot be
stabilized against the growth of bending perturbations of any wavelengths.
The theory constructed for a homogeneous infinitely thin layer does not
yield this regime. It should be borne in mind, however, that the
contribution of the destabilizing term ($\nu_{\rm d}^2$) must decrease with
increasing disk thickness; therefore, the instability will be saturated at a
large, but finite disk thickness. However, the following might be expected:
other things being equal, the hotter the initial model in the plane, i.e.,
the larger the Toomre parameter $Q_{\rm T}$, the higher the saturation level.

\section{BENDING INSTABILITY: NUMERICAL SIMULATIONS OF THREE-DIMENSIONAL
DISKS}

The nonlinear growth stages of bending instability in inhomogeneous
finite-thickness stellar disks have been extensively investigated by
numerically solving the gravitational $N$-body problem for various stellar
disk models. Raha {\it et al.} (1991) first observed the bending instability
of bars in their numerical simulations; Sellwood and Merritt (1996) and
Merritt and Sellwood (1994) studied the nonlinear regime of bending
instability in nonrotating disks with the radial density profiles that
corresponded to the Kuzmin--Toomre model (see, e.g., Binney and Tremaine
(1987), p. 43); Griv {\it et al.} (1998) numerically analyzed the
development of a bend in a layer of newly formed stars; Tseng (2000)
simulated the evolution of the vertical structure of a homogeneous,
initially thin disk of finite radius; Sotnikova and Rodionov (2003)
considered a rotating disk with an exponential density profile along the $R$
axis and assumed the presence of a dark halo in the 
system. The latter authors analyzed the question of how the evolution of
initially equilibrium thin disks depends on the governing parameters of the
bending instability, which include the initial disk half-thickness $z_0$,
the Toomre parameter $Q_{\rm T}$, and the relative mass of the dark halo
$M_{\rm h} / M_{\rm d}$ within a fixed radius. Below, we list the most
important conclusions that follow from the $N$-body simulations described in
the literature. These conclusions in many respects agree with those given in
the section entitled ``Peculiarities of the growth of bending
instability{\ldots}'' of this paper for inhomogeneous infinitely thin disks.

(1) In inhomogeneous models, all of the experimentally observed modes are
global, i.e., the scale length of the unstable perturbations is larger than
the disk scale. The linear theory constructed for homogeneous models yields
no such result.

(2) The saturation level for the bending instability depends on $\sigma_R$
(or $Q_{\rm T}$). The larger the reserve of disk strength against
perturbations in the disk plane, the greater the dificulty to stabilize the
disk against the growth of bending perturbations. A rapid and significant
(occasionally catastrophic) increase in the disk thickness, particularly in
the central regions, to values that are severalfold larger than those
yielded by a linear analysis for homogeneous, moderately hot models (Toomre
1966; Kulsrud {\it et al.} 1971; Polyachenko and Shukhman 1977; Araki 1985)
was observed in all of the simulations with initially hot disks (Sellwood and 
Merritt 1994; Tseng 2000; Sotnikova and Rodionov 2003). This mechanism may
be responsible for the formation of central bulges in spiral galaxies, at
least it can feed the spherical component with new objects.

(3) The central regions of the disk are most unstable (Sellwood and Merritt
1994; Merritt and Sellwood 1994; Griv and Chiueh 1998; Griv {\it et al.} 2002;
Sotnikova and Rodionov 2003). It is here that the bending modes are formed.
Subsequently, their amplitude increases, sometimes significantly. This is
particularly true for the perturbations with $m = 0$. The nonaxisymmetric
modes (with the azimuthal numbers $m = 1$ and $m = 2$) drift to the disk
periphery, temporarily creating the effect of a largescale warp of the
entire galaxy, and are then damped\footnote{In contrast to the result of
Griv {\it et al.} (2002), the largescale S-shaped or U-shaped warp of the
galactic edge always disappeared on long time scales (> 5 Gyr) in all of the
simulations by Sotnikova and Rodionov (2003).}.
The instability in the central regions of the stellar disks is saturated at
(Sotnikova and Rodionov 2003)
$$
\sigma_z / \sigma_R \approx 0.75 \, - \, 0.8 \, .
$$

(4) The presence of a massive dark halo, which was included in the models by
Sotnikova and Rodionov (2003), has always been a stabilizing factor that
suppresses the growth of bending modes. This effect appears to have been
first described qualitatively by Zasov {\it et al.} (1991).

\section{THE STABILIZING EFFECT OF THE SPHEROIDAL COMPONENT: A QUALITATIVE
ANALYSIS}

The remarkable agreement (at least on a qualitative level) between the
conclusions that follow from the analysis of the dispersion relation for an
infinitely thin disk and the results of numerical simulations with
three-dimensional disks allows us to analyze the applicability of yet
another conclusion that can be drawn from Eq.~(\ref{dispersion2}) to
finite-thickness disks. Since the central regions of the disk are most
unstable, it is important to separate out the factors that have a
stabilizing effect precisely on these regions. An additional spherical
component (a dark halo, a bulge) can be such a stabilizing factor. In this
case, another term related to the restoring force exerted from the spherical
component appears in Eq.~(\ref{dispersion2}), with
\be
\nu^2_{\rm sph} =
\left.\frac{\partial^2 \Phi_{\rm sph}(r)}{\partial z^2}\right|_{z=0} > 0
\, .
\ee

This term was also introduced by Sellwood (1996) when analyzing the
dispersion relation for the longwavelength bending perturbations of an
infinitely thin disk, but its role was not studied.

Zasov {\it et al.} (1991, 2002) and Mikhailova {\it et al.} (2001) concluded
that the minimum possible relative thickness of an equilibrium stellar disk,
$z_0/h$, decreases with increasing relative mass of the dark halo. Let us
consider the relationship between the stellar disk thickness and the mass of
the spheroidal component in terms of the stabilization conditions for the
bending modes in inhomogeneous thin disks.

We take specific bulge and halo models (these models were subsequently used
in our numerical calculations). A Plummer sphere is taken as the bulge
model. Its potential is (see, e.g., Binney and Tremaine (1987), pp. 42--43)
\be
\label{eq_bulge}
\Phi_{\rm b}(r) = -
\frac{G M_{\rm b}}{\left(r^2 + a_{\rm b}^2\right)^{1/2}} \, ,
\ee
where $M_{\rm b}$ is the total mass of the bulge, and $a_{\rm b}$ is the
scale length of the matter distribution. We describe the potential of the
dark halo in terms of the logarithmic potential (see, e.g., Binney and
Tremaine (1987), p. 46)
\be
\label{eq_halo}
\Phi_{\rm h}(r) = \frac{v_{\infty}^2}{2}\ln(r^2+a_{\rm h}^2) + const \, ,
\ee
where $a_{\rm h}$ is the scale length, and $v_{\infty}$ is the velocity of a
particle in a circular orbit of infinite radius. The parameter $v_{\infty}$
is related to the mass of the halo with a sphere of given radius $r$ by 
$M_{\rm h}(r) =
\displaystyle \frac{v_{\infty}^2}{G} \frac{r^3}{r^2 + a_{\rm h}^2}$).

For the additional stabilizing term in the dispersion relation, the models
of the spherical components (the bulge and the halo) that we used yield
$$
\nu^2_{\rm b} = \displaystyle\frac{G M_{\rm b}}{(R^2+a_{\rm b}^2)^{3/2}}
\, ; \,\,\,\,\,\,\,\,\,\,
\nu^2_{\rm h} = \displaystyle\frac{v_{\infty}^2}{R^2+a_{\rm h}^2} \, .
$$
The stabilizing effect of the spherical component weakens at large $R$, but
the disk itself in the peripheral regions has a stabilizing effect:       
$F_z \approx - G M_{\rm d} z/R^3$ at 
$|z| << R \Rightarrow
\nu_{\rm d}^2 = - \partial F_z / \partial z > 0$. On the other hand, the
strength of the effect increases in the central (most unstable) regions
(i.e., for $R \to 0$); this strength depends not only on the total mass of
the spherical component ($M_{\rm b}$ or $v_{\infty}$), but also on the
degree of matter concentration toward the center, $a_{\rm b}$ and 
$a_{\rm h}$. It thus follows that the presence of a compact (not necessarily
massive) bulge in galaxies may prove to be enough to suppress the bending
perturbations. This implies that the disks of galaxies with compact bulges
can be as thin as the disks embedded in a massive dark halo.

To test our conclusion, we carried out a series of numerical simulations.

\section{THE STABILIZATION OF BENDING PERTURBATIONS BY A COMPACT BULGE:
$N$-BODY SIMULATIONS}

\subsection{The Method}

We used an algorithm based on the hierarchical tree construction method
(Barnes and Hut 1986) to simulate the evolution of a self-gravitating
stellar disk. In our calculations, we always included the quadrupole term in
the Laplace expansion of the potential produced by groups of distant bodies.
The parameter $\theta$ (Hernquist 1993) that is responsible for the accuracy
of calculating the gravitational force was chosen to be $0.7$ in all our
simulations. The NEMO software package (http://astro.udm.edu/nemo; Teuben
1995) was taken as the basis. We enhanced the capabilities of this package
by including several original codes for specifying the equilibrium initial
conditions in a flat stellar 
system\footnote{The mkexphot code for specifying equilibrium stellar disk
models from the NEMO package has limitations on the parameters of the outer halo and does not enable the gravitational
field of the bulge to be specified. The models constructed using our codes
closely agree with those obtained using the mkexphot code for identical
parameters.} and supplemented it with new codes that allow us to easily
analyze the data obtained and to present them in convenient graphical and
video formats.

\subsection{The Numerical Model}

When constructing the galaxy model, we distinguished two components in it: a self-gravitating stellar
disk and a spherically symmetric component that was described in terms of
the external static potential, which is a superposition of two potentials,
(\ref{eq_bulge})~for the bulge and (\ref{eq_halo})~for the dark halo. At
large distances $R$ from the center of the stellar system, in the region
where the halo dominates, potential~(\ref{eq_halo}) yields a flat rotation
curve. The disk was represented by a system of $N$ gravitating bodies with
the density profile~(\ref{eq_star_disk_dens}).

The initial conditions in the $N$-body problem suggest specifying the mass,
position in space, and three velocity components for each particle. The
particle coordinates are naturally determined in accordance with the disk
matter density profile~(\ref{eq_star_disk_dens}); the distant regions of the
disk are disregarded. We took only those particles for which the cylindrical
radius $R < R_{\rm max}$ and $|z| < z_{\rm max}$. The mass of all particles
was assumed to be the same. The total mass of the particles was equal to the
mass of the disk region under consideration (i.e., the disk region for which
$R < R_{\rm max}$ and $|z| < z_{\rm max}$). The particle velocities were
specified using the equilibrium Jeans equations by the standard technique
(see, e.g., Hernquist (1993), Section 2.2.3).

\subsection{Specifying the Velocity Field in the Model Galaxy}

To specify the initial particle velocities for a disk that is in equilibrium
in the plane and in the vertical direction, we make the following
assumptions:

(1) The velocity distribution function is the Schwarzschild one; in other
words, the particle velocity distribution function has only four nonzero
moments: the mean azimuthal velocity $\bar v_{\varphi}$, the radial
velocity dispersion $\sigma_R$, the azimuthal velocity dispersion
$\sigma_{\varphi}$, and the vertical velocity dispersion\footnote{As many
members of the astronomical society, we have the
bad habit of calling the standard of the distribution function
the dispersion.} $\sigma_z$.

(2) All four moments depend only on the cylindrical
radius $R$ and do not depend on $z$.

(3) The epicyclic approximation is valid\footnote{In the central regions of
the disk, this approximation breaksdown.}.

(4) $\sigma_R^2$ is proportional to the surface density of the
stellar disk, i.e., $\sigma_R \propto \exp{(- R/2h)}$ (this assumption
agrees well with the observational data; see, e.g., van der Kruit and Searle
1981).

The following relations for the moments (in which the $R$ dependence was
omitted for simplicity) can then be derived from the Jeans equations (see,
e.g., Binney and Tremaine 1987):
\be
\label{eq_init_data_01}
\left\{
\begin{array}{rcl}
{\bar v}_{\varphi}^2 &=& v_{\rm c}^2 + \sigma_R^2 - \sigma_{\varphi}^2 +
\displaystyle \frac{R}{\Sigma_{\rm d}}
\frac{\partial \Sigma_{\rm d} \sigma_R^2}{\partial R} \, , \\
\sigma_{\varphi} &=&
\sigma_R \, \displaystyle\frac{\kappa}{2 \Omega} \, , \\
\sigma_z^2 &=&
\pi \Sigma_{\rm d} z_0 \, ,\\
\end{array}
\right.
\ee
where $v_{\rm c}$ is the circular velocity of a particle placed in the total
potential of the disk and the spherical component, the bulge and the halo,
($v_{\rm c}^2 = v_{\rm c,d}^2 + v_{\rm c,b}^2 + v_{\rm c,h}^2$;
$\displaystyle v_{\rm c,b}^2 = R \frac{\partial \Phi_{\rm b}}{\partial R}$;
$\displaystyle v_{\rm c,h}^2 =
R \frac{\partial \Phi_{\rm h}}{\partial R}$), 
$\displaystyle \Omega = \frac{v_{\rm c}}{R}$ is the angular velocity, and 
$\displaystyle \kappa = \sqrt{ 2 \frac{v_{\rm c}^2}{R^2} +
\frac{1}{R} \, \frac{dv_{\rm c}^2}{dR}}$ is the
epicyclic frequency.

The circular velocities for the bulge~(\ref{eq_bulge}) and the
halo~(\ref{eq_halo}) have analytical expressions. The circular velocity
for the disk~(\ref{eq_star_disk_dens}) can be determined by numerical
integration (see the section entitled ``Global modes'') using the general
formula 
\be
v_{\rm c,d}^2({\bf r}) =  \int G
\rho_{\rm d}({\bf r^{\prime}}) \cdot
\frac{({\bf r}^{\prime} - {\bf r})
\cdot {\bf R}}{|{\bf r}^{\prime} - {\bf r}|^3} \, \,
d^3 {\bf r}^{\prime}  \, ,
\label{v_disk}
\ee
where {\bf R} is the projection of the vector {\bf r} onto the disk plane.

The Jeans equations that are used to derive
relations~(\ref{eq_init_data_01}) are known to provide no exact disk
equilibrium (see, e.g., Binney and Tremaine 1987). Moreover, the last
relation in~(\ref{eq_init_data_01}), which follows from the vertical
equilibrium condition for a disk with the density
profile~(\ref{eq_star_disk_dens}) and a $z$-independent $\sigma_z$, was
written without including the influence of the additional spheroidal
components. The adjustment to equilibrium occurs on time scales of the order
of several vertical oscillation times $1/\nu_{\rm d}$. This time was always
no longer than $100-120$ integration time steps for the equations of motion
(the thinner the disk, the shorter this time) and much shorter than the
instability growth time scale in the disk. Moreover, in the context of the
problem of the growth and saturation of unstable modes, a small deviation of
the disk from equilibrium at the initial time may be treated as an
additional initial perturbation.

The fourth assumption (see above) about the radial velocity dispersion
$\sigma_R^2(R)$ causes dificulties in calculating
${\bar v}_{\varphi}$ in the central regions. In the first equation
of system~\ref{eq_init_data_01}, ${\bar v}_{\varphi}^2$ is occasionally
negative at small $R$ (since $\sigma_R^2(R)$ rapidly increases toward the
center, the last term on the right-hand side can make a large negative
contribution). For this reason, the dependence for $\sigma_R$ was reduced at
the center (Hernquist 1993):
\be
\label{eq_soft_sigma_R}
\sigma_R \propto \exp\left(-\sqrt{R^2 - 2a_s^2}/2h\right) \, .
\ee

If the parameter $a_s$ is taken to be $h/4 - h/2$, then this proves to be
enough to properly calculate ${\bar v}_{\varphi}$. In the central regions of
the disk, $\sigma_z$ adjusted to $\sigma_R$ in such a way that the ratio
$\sigma_z / \sigma_R$ was constant at a given half-thickness $z_0 = const$
in the initial model throughout the disk.

The proportionality factor in~(\ref{eq_soft_sigma_R}) can be determined via
the Toomre parameter $Q_{\rm T}$ at some radius $R_{\rm ref}$:
\be
\label{toomre}
\sigma_R(R_{\rm ref})
= Q_{\rm T} \, \sigma_R^{\rm cr}(R_{\rm ref})
= Q_{\rm T}
\frac{3.36 \cdot \Sigma_{\rm d}(R_{\rm ref})}{\kappa(R_{\rm ref})}
\, .
\ee
The sought proportionality factor can be obtained
from~(\ref{eq_soft_sigma_R}) and~(\ref{toomre}). Specifying $Q_{\rm T}$,
which ensures disk stability in the plane, at $R_{\rm ref} \approx 2.5 h$
yields the condition $Q_{\rm T}(R) \geq Q_{\rm T}(R_{\rm ref})$ 
(Hernquist 1993). The latter, in turn, ensures a stability level against
perturbations in the disk plane no lower than that at $R_{\rm ref}$. The
initial half-thickness $z_0$ for the adopted $Q_{\rm T}(R_{\rm ref})$ was
chosen in such a way that ?z/?R was less than $0.3-0.4$, which ensured
initial instability against the growth of bending modes.

\subsection{Parameters of the Problem}

All of the results discussed below are presented in the following system of
units: the gravitational constant is $G = 1$, the unit of length is 
$R_u = 1$~kpc, and the unit of time is $t_u = 1$~Myr. The unit of mass is
then $M_u = R_u^3 / G t_u^2 = 22.2 \cdot 10^{10} M_{\odot}$, and the unit of
velocity is $v_u = R_u / t_u = 978$~km s$^{-1}$.

The number of bodies in the simulations was $N = 300\,000$ (in several
cases, $500\,000$ and $600\,000$). The force of interaction between two
particles with coordinates ${\bf r}_i$ and ${\bf r}_j$ and masses $m_i$ and
$m_j$ was modified, as is commonly done, as follows:
$$
{\bf F}_{ij} =
G m_i m_j\frac{{\bf r}_j -
{\bf r}_i}{(|{\bf r}_j - {\bf r}_i|^2 + \epsilon^2)^{3/2}} \, ,
$$
where $\epsilon$ is the softening length of the potential produced by an
individual particle. When collisionless systems are simulated, this
parameter is introduced for two reasons. First, the divergence of the
interaction force in close particle--particle encounters must be avoided
when integrating the equations of motion. Second, when the phase density of
a collisionless system is represented by a finite number of particles, the
inevitable fluctuations in the particle distribution must be smoothed in
such a way that the forces acting in the system being simulated were are to
the forces acting between the particles in a system with a smoother density
profile. The softening length $\epsilon$ was chosen to be
$0.02$. This value is approximately a factor of 2 or 3 smaller than the mean
separation between the particles (at $N = 300\,000$) within the region
containing half of the disk mass. On the one hand, it matches the criterion
for choosing $\epsilon$ based on minimization of the mean irregular force
(Merritt 1996) and, on the other hand, allows the vertical structure of thin
disks to be adequately resolved.

To integrate the equations of motion for particles in the self-consistent
potential of the disk and the external field produced by the spheroidal
component, we used a leapfrog scheme that ensured the second order of
accuracy in time step. The time step was $0.5$ (in several models,
$0.25$)\footnote{The choice of the time step is limited above by
$\epsilon$ --- the particle must take at least one step on the smoothing
length.}.

We constructed a total of about 60 models. The entire set of models can be
arbitrarily divided into two classes: the models with and without bulges.
The scale length of the density distribution in the bulge $a_{\rm b}$ was
assumed to be equal to $0.5$ almost for all of the models with bulges. In
several models without bulges, we chose a concentrated halo 
($a_{\rm h} = 2$). In all of the remaining cases, the halo was ``looser''
($a_{\rm h} = 10$). The total relative mass of the spheroidal components
$\mu = M_{\rm sph}(4h) / M_{\rm d}(4h)$ was varied over the range 0.25
to 3.5

The disk in our models has the following parameters\footnote{These
parameters are close those of the disk in our Galaxy.}: $h = 3.5$ and the
disk mass (in dimensional units) 
$M_{\rm d}(4h)= (4-8) \times 10^{10} M_{\odot}$. The initial thickness was
varied over the range $z_0 = 0.1 - 0.5$. In order not to abruptly cut off
the model disk at the radius corresponding to the optical radius ($\sim 4h$),
we chose $R_{\rm max} = 25$, with $z_{\rm max} = 5$. The smoothing
parameter of the initial radial profile of the velocity dispersion
$\sigma_R$ is $a_s = 1$. The parameter $Q_{\rm T}$ in the
discussion of our simulations is given for the radius $R_{\rm ref} = 8.5$.

\subsection{Simulation Results}

Previously (Sotnikova and Rodionov 2003), we showed that there are two
distinct vertical stellar disk relaxation mechanisms related to bending
instability: the bending instability of the entire disk and the bending
instability of the bar forming in the disk. The former mechanism dominated
in galaxies that are hot in the plane ($Q_{\rm T} \gsim 2.0$), and the bar
formation was suppressed in this case; the latter mechanism dominated in
galaxies with a moderate Toomre parameter $Q_{\rm T}$ (such galaxies were
unstable against the growth of a bar mode). In the simulations whose results
are presented and analyzed below, we also considered two distinct cases: hot
disks ($Q_{\rm T} = 2.0$) in which only bending instability developed, and
cooler models ($Q_{\rm T} = 1.5$) --- here, we observed the combined effect
of the two types of instability.

{\bf Hot disks.} For hot (in the plane) disks ($Q_{\rm T} = 2.0$),
we revealed distinct patterns of growth and saturation of bending
perturbations that are consistent with the conclusions following from a
qualitative analysis of the dispersion relation~(\ref{dispersion2}).

\begin{figure}
\centerline{\psfig{file=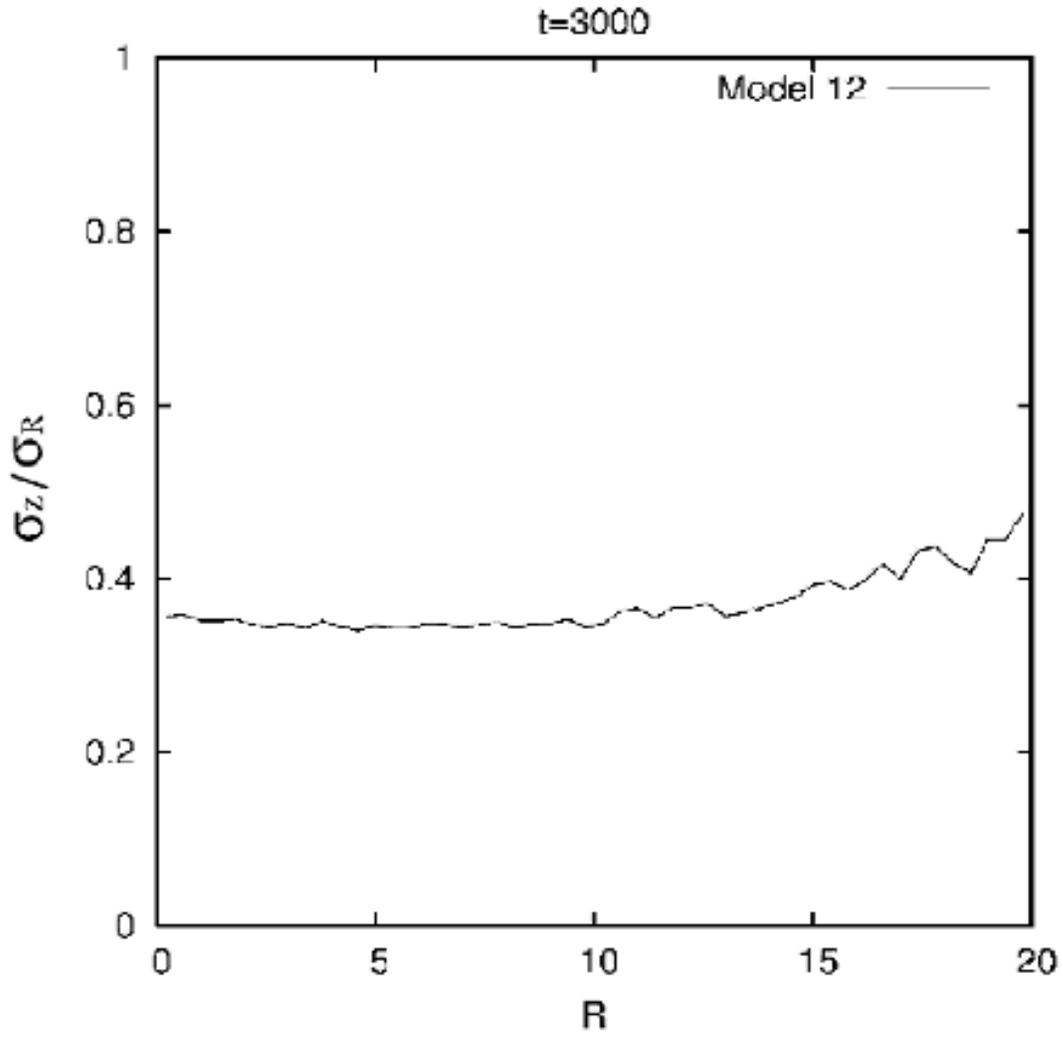,width=16cm}}
\caption[2]{Ratio $\sigma_z / \sigma_R$ versus $R$ for the time 
$t = 3000$
(model~12 with a massive halo: $\mu = 3.0$, $\mu_{\rm b} = 0$).}
\label{big_halo}
\end{figure}

The stabilizing effect from the presence of a massive spherical component
that was discussed in the section entitled ``The stabilizing effect of the
spheroidal component{\ldots}`` is clearly seen in Fig.~\ref{big_halo}.
This figure shows the radial profile of $\sigma_z / \sigma_R$ for
model~12 with $M_{\rm b} = 0$ and $\mu = 3.0$. Throughout the disk,
$\sigma_z / \sigma_R$ was set at $\approx 0.35$ and was determined not by
the bending instability, but by the disk heating through the scattering of
stars by inhomogeneities related to the different disk thickness in the
spiral arms and in the interarm space (Sotnikova and Rodionov 2003).

\begin{figure}
\centerline{\psfig{file=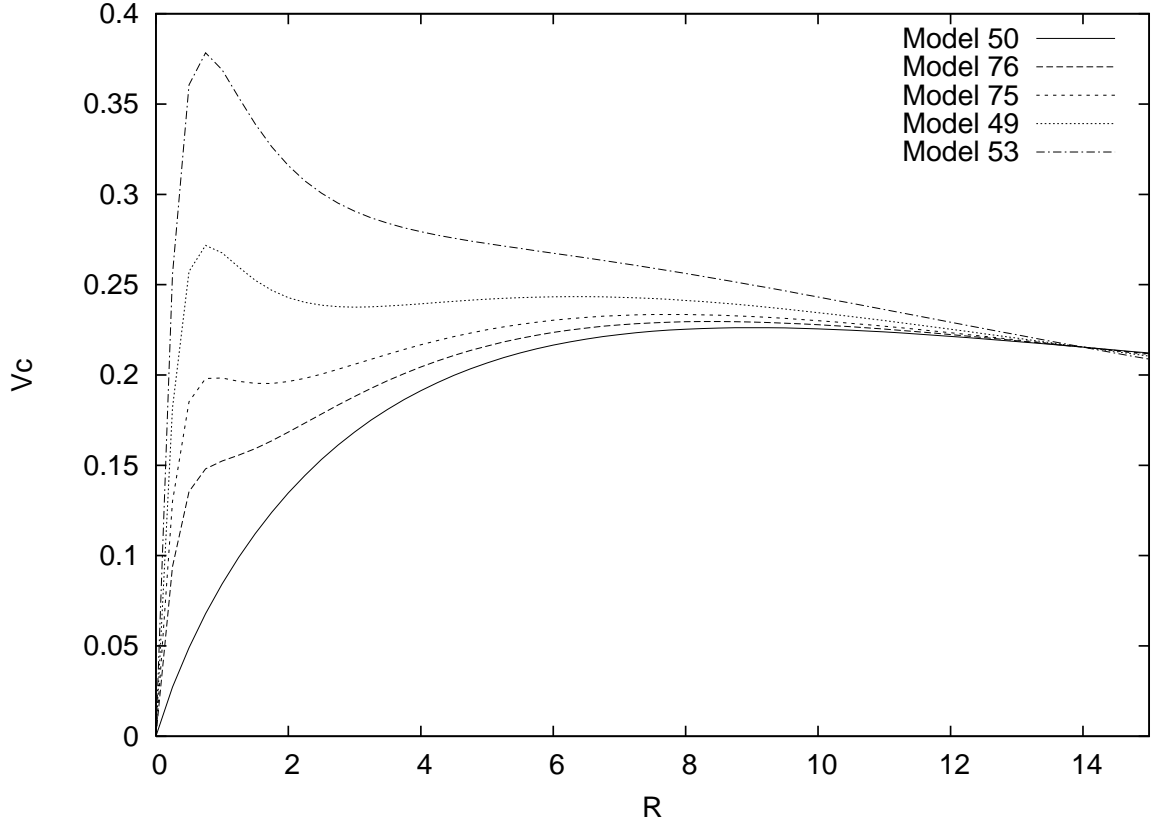,width=16cm,angle=270}}
\caption[3]{Initial rotation curves for models~50, 76, 75, 49, 
and~53. The ratio
of the total mass of the spherical component to the mass of the disk within
a radius of $4h$ is the same for all models, $\mu = 0.5$; $\mu_{\rm b} = 0$
for model~50, $\mu_{\rm b} = 0.0625$ for model~76, $\mu_{\rm b} = 0.125$ for
model~75, $\mu_{\rm b} = 0.25$ for model~49, and $\mu_{\rm b} = 0.5$ for
model~53. The unit of velocity is $978$~km s$^{-1}$.}
\label{vcirc}
\end{figure}

We will demonstrate the stabilizing effect of a compact (not necessarily
massive) bulge comparable to the effect of a massive dark halo with a broader
density profile using the results obtained for the following group of models
as an example: 50, 76, 75, 49, and 53. In all five models, the total mass of
the spheroidal component is the same and accounts for half of the diskmass
within four exponential disk scale lengths, $\mu = 0.5$, but it is
differently distributed between the two spherical subcomponents. In model~50,
all of the mass is contained in the halo 
($\mu = \mu_{\rm h} = M_{\rm h}(4h) / M_{\rm d}(4h)$); in model~53, only a
compact bulge is present\footnote{In model~53, a bulge with a mass equal
to half of the disk mass and a scale length of 500~pc is atypical of real
galaxies. We consider this as a limiting case.} 
($\mu = \mu_{\rm b} = M_{\rm b}(4h) / M_{\rm d}(4h)$). The remaining
models are intermediate between the two extreme models. The initial
thickness for all of the models was chosen to be the same, $z_0 = 0.1$. The
rotation curves for these models are shown in Fig.~\ref{vcirc}. The variety
of the shown curves to some extent reflects the actual variety of rotation
curves for spiral galaxies.

\begin{figure}
\centerline{\psfig{file=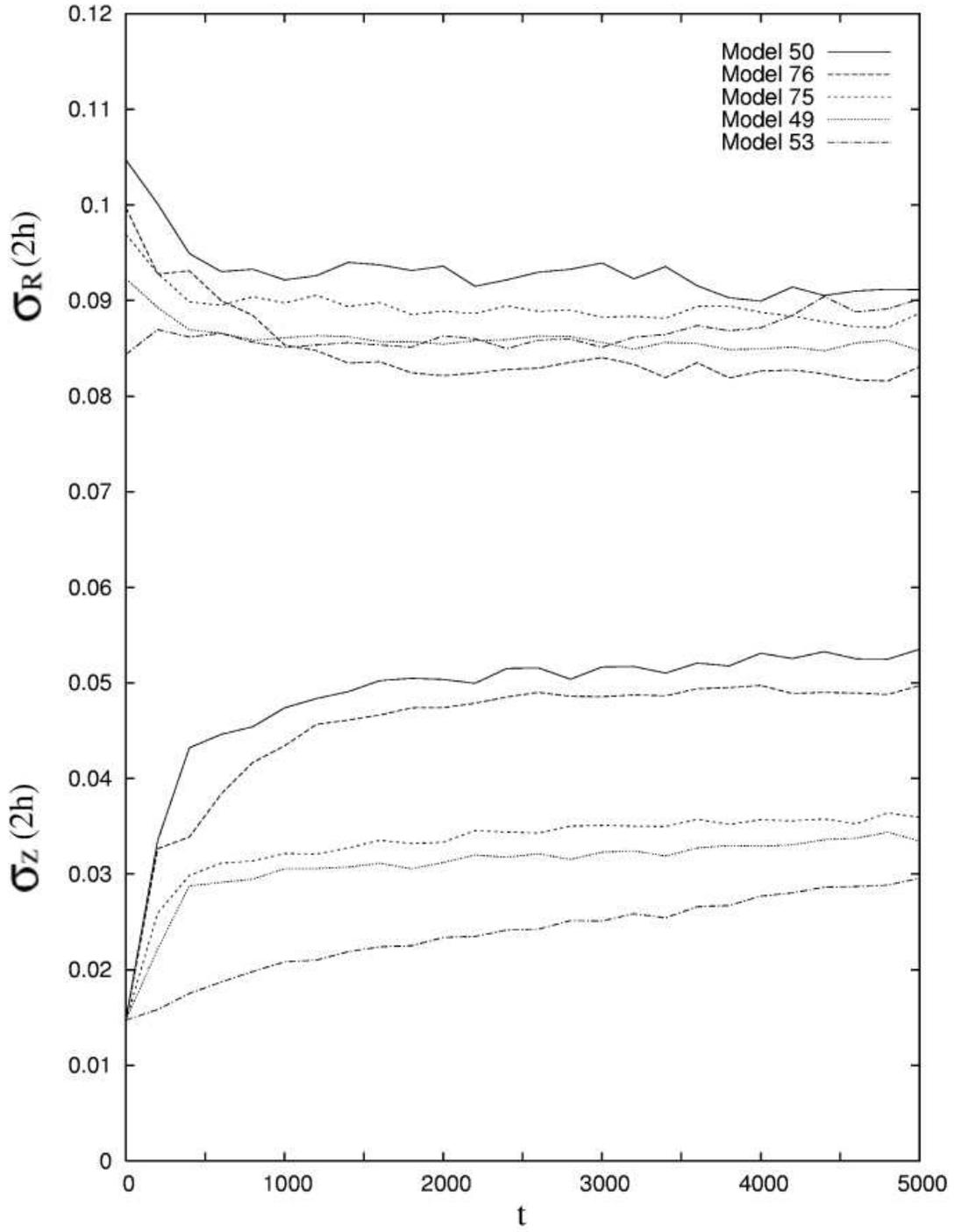,width=16cm}}
\caption[4]{Evolution of the velocity dispersions $\sigma_R$ and
$\sigma_z$ at
$R = 2h$ for the same models as those in Fig.~\ref{vcirc}.}
\label{sigma_z_R_t}
\end{figure}

We traced the evolution of these models up to $t = 5000$.
Figure~\ref{sigma_z_R_t} illustrates the variations in the dynamical
parameters of the disk $\sigma_R$ and $\sigma_z$ the radial and vertical
velocity dispersions calculated at $R = 2h$. All of the models demonstrate
an initial increase in $\sigma_z$ and a decrease in $\sigma_R$. Subsequently
(after $t \approx 1000$), the latter parameter reaches an approximately
constant value, while $\sigma_z$ for some of the models (this is primarily
true for the model with a massive bulge) continues to slowly increase. The
number of particles in our models and the softening length were chosen in
such a way that the two-body relaxation time was much longer than the time
scale on which we considered the evolution of our numerical models. The
absence of heating related to numerical relaxation is confirmed by the
behavior of $\sigma_R$ and the preservation of the pattern of evolution of
the system as the number of particles increases to $N = 600 000$. The
continuing small secular increase in the vertical velocity dispersion
probably reflects the fact that some of our models did not reach a steady
state\footnote{If the system has no third integral of motion, then its
evolution to equilibrium must eventually lead to the relation 
$\sigma_z = \sigma_R$.}.

\begin{figure}
\centerline{\psfig{file=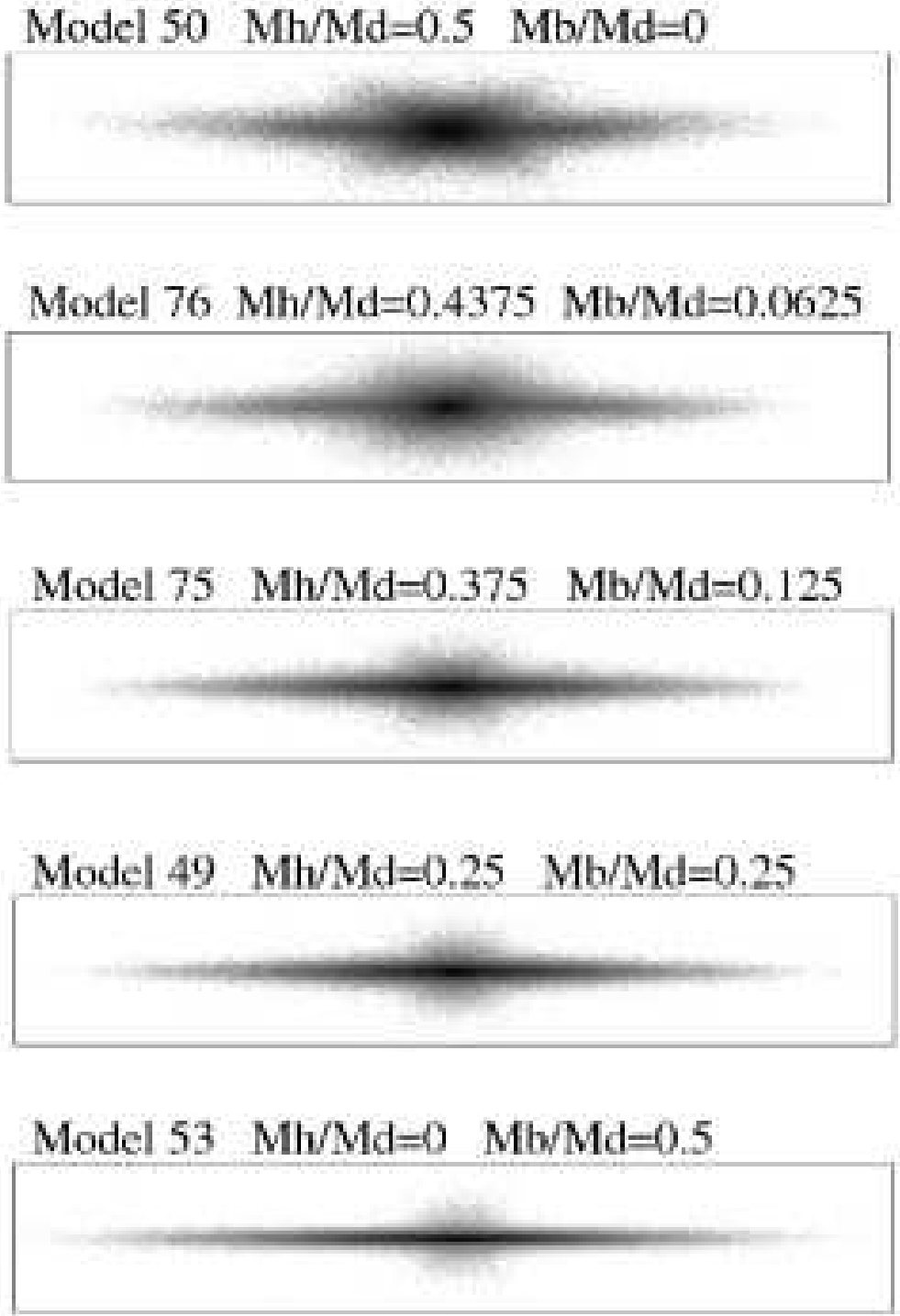,width=14cm}}
\caption[5]{Edge-on view of the galaxy at the time $t = 3000$ for
models~50, 76,
75, 49, and~53. The strenght of the image blackening corresponds
to the logarithm of the particle number per pixel. The horizontal and
vertical scales are 60 and 10, respectively. The ratio of the
total mass of the spherical component to the mass of the disk within a
radius of $4h$ is the same for all models, $\mu =0.5$.}
\label{edge_on}
\end{figure}

Figure~~\ref{edge_on} shows five frames that correspond to the late
evolutionary stages of our model disks. As expected, the saturation level
for the bending instability in model 50 was very high and did not match the
standard linear criterion. The galaxy greatly thickened at the final
evolutionary stages. However, when we transferred 50\% of the mass from the
halo to the bulge (model~49: $\mu_{\rm h} = \mu_{\rm b} = 0.25$), the
picture changed. The saturation level for the bending instability became
much lower. At the final evolutionary stages, the disk was much thinner than
that in model~50. In model~53, when we placed all of the mass of the
spherical component in the bulge, the amplitude of the observed bend was
very low, and the galaxy remained quite thin even at the late evolutionary
stages.

The disk thickness can be quantitatively estimated as the 
root-mean-square~(rms) value of the $z$ coordinates of the disk particles, 
$\displaystyle z_{rms} = \sqrt{<z^2> - <z>^2}$. This estimate is commonly
encountered in the literature. It can be shown that for the vertical density 
profile~(\ref{eq_star_disk_dens}), the relationship between this parameter
and $z_0$ is given by $z_{rms} = \pi / 2 \sqrt{3} z_0 \approx 0.91 z_0$. In
practice, however, this parameter proved to be a not very good
characteristic of the thickness. First, the fluctuations in this parameter
along $R$ were found to be great even when using a large number of particles
if only no averaging is performed in concentric rings of large width.
Second, the thickness calculated in this way turns out to be systematically
overestimated due to the existing of a significant tail of the particles that
went far from the disk plane. For these reasons, we estimated the disk
thickness at a given distance $R$ through the median of the absolute value
of $z$ that was designated as $z_{1/2}$. Twice the value of $z_{1/2}$ is
nothing but the disk thickness within which half of the particles is
contained. For the density profile~(\ref{eq_star_disk_dens}), 
$2z_{1/2} = z_0 \cdot \ln{3} \approx 1.1 z_0 $.

\begin{figure}
\centerline{\psfig{file=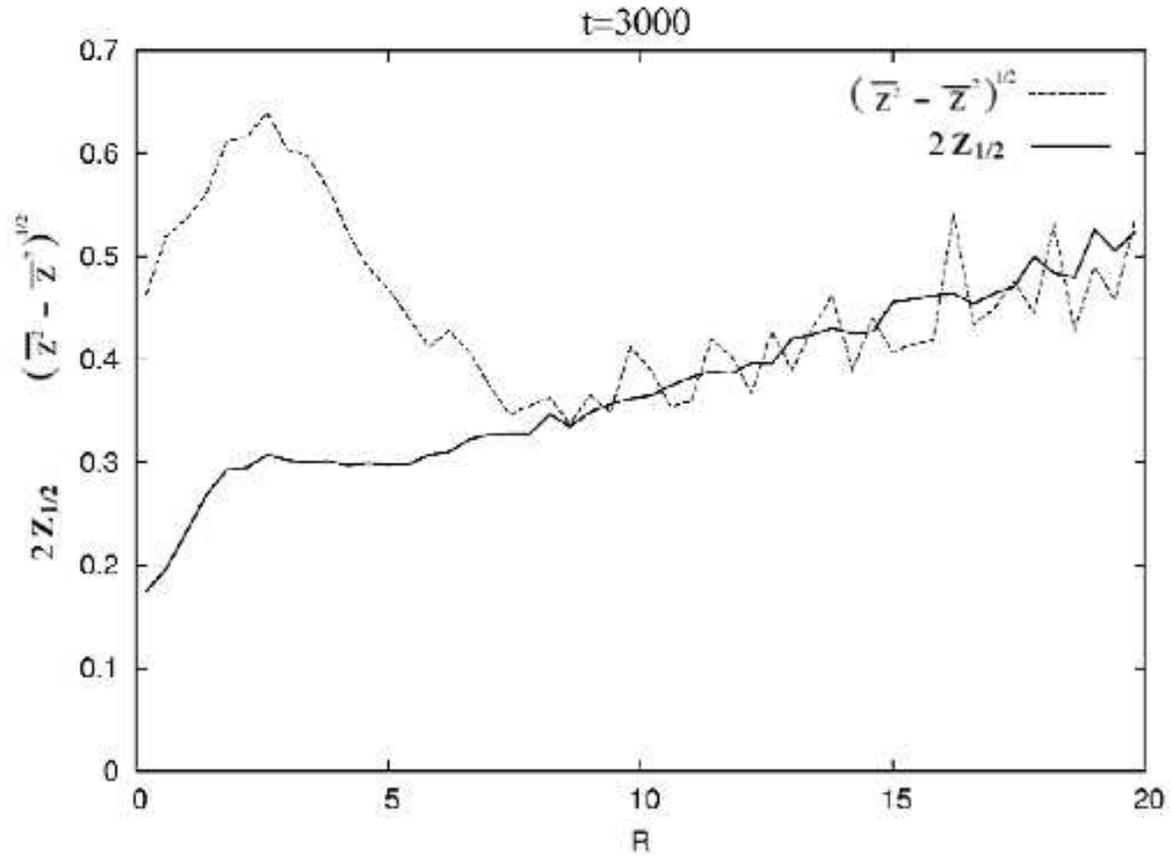,width=16cm}}
\caption[6]{Radial thickness profiles for the galaxy at the time 
$t = 3000$ for
model~53. The thickness was determined by two methods: as the rms value of
the $z$ coordinates of the disk particles, $z_{rms}$, and as twice the median
of $|z|$, $2z_{1/2}$.}
\label{thickness}
\end{figure}

Figure~\ref{thickness} shows the differences between the radial disk
thickness profiles for model~53 obtained by the two described methods (the
averaging was performed in concentric rings; the ring width was 
$\Delta R = 0.4$). We see that $z_{1/2}$ behaves much more smoothly (we have
in mind the overall monotonic dependence of the density and the fluctuation
level) than does the rms value of the $z$ coordinate commonly used to
estimate the disk thickness in $N$-body simulations.

\begin{figure}
\centerline{\psfig{file=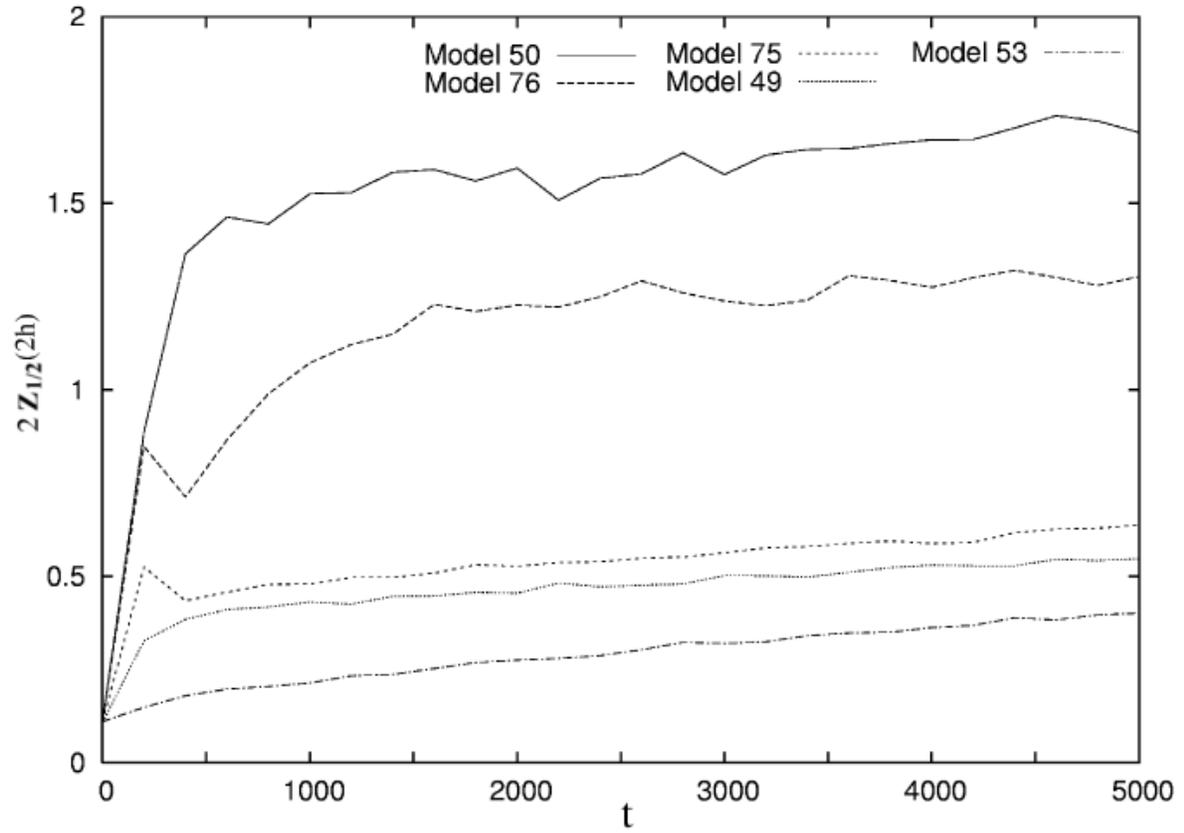,width=16cm}}
\caption[7]{Radial thickness profiles for the galaxy ($2z_{1/2}$) at
the time
$t = 3000$ for models~50, 76, 75, 49, and~53.}
\label{median_z_R}
\end{figure}

\begin{figure}
\centerline{\psfig{file=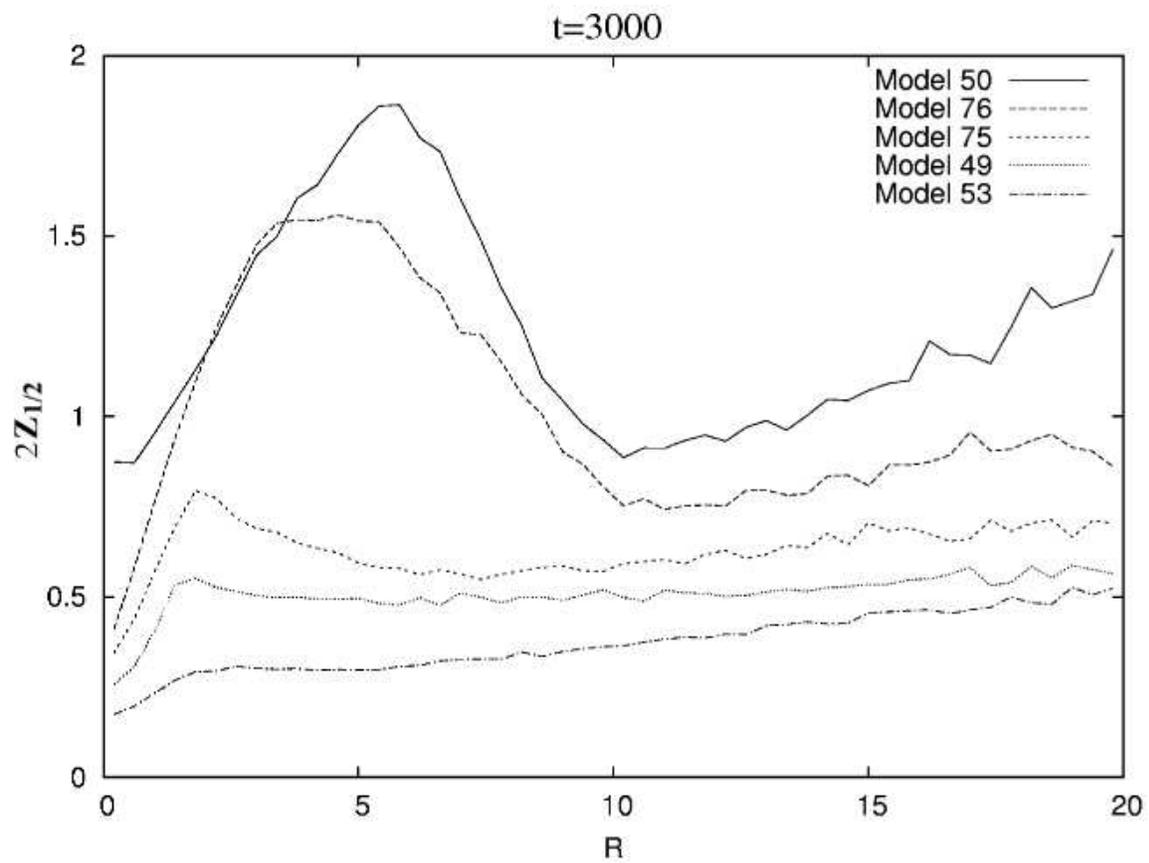,width=16cm}}
\caption[8]{Evolution of the disk thickness ($2z_{1/2}$) at $R = 2h$
for
models~50, 76, 75, 49, and~53 (see the caption to Fig.~\ref{vcirc}).}
\label{median_z_t}
\end{figure}

Figure~\ref{median_z_R} shows the radial disk thickness profile for
models~50, 76, 75, 49, and 53 at the time $t = 3000$. Note that the
thickness profile for model 50 is rather unusual in shape. This shape is
most likely attributable to the existence of X-shaped stationary orbits in
the central regions that arise at a certain disk thickness when there are
conditions for the resonance between the stellar oscillation frequencies in
the disk plane and in the vertical direction (see, e.g., 
Patsis {\it et al.} 2002). We see the following from Fig.~\ref{median_z_R}, as
well as from Fig.~\ref{edge_on}, which show the edge-on views of the model
galaxies: the thinner the galactic disk, the larger the mass of the
spheroidal component contained in a compact bulge. Since not all of our
models reached a steady state, their thickness continues to slowly increase
(Fig.~\ref{median_z_t}), but the differences in thickness are always
preserved. Similar results were obtained for all of the remaining such
models with the same mass of the spherical component in the range
$M_{\rm sph}(4h) = 0.25 M_{\rm d}(4h)$ to 
$M_{\rm sph}(4h) = 3.5 M_{\rm d}(4h)$. The stabilizing effect of a bulge was
particularly pronounced in those cases where the bulk of the galactic mass
was contained in the disk. 

\begin{figure}
\centerline{\psfig{file=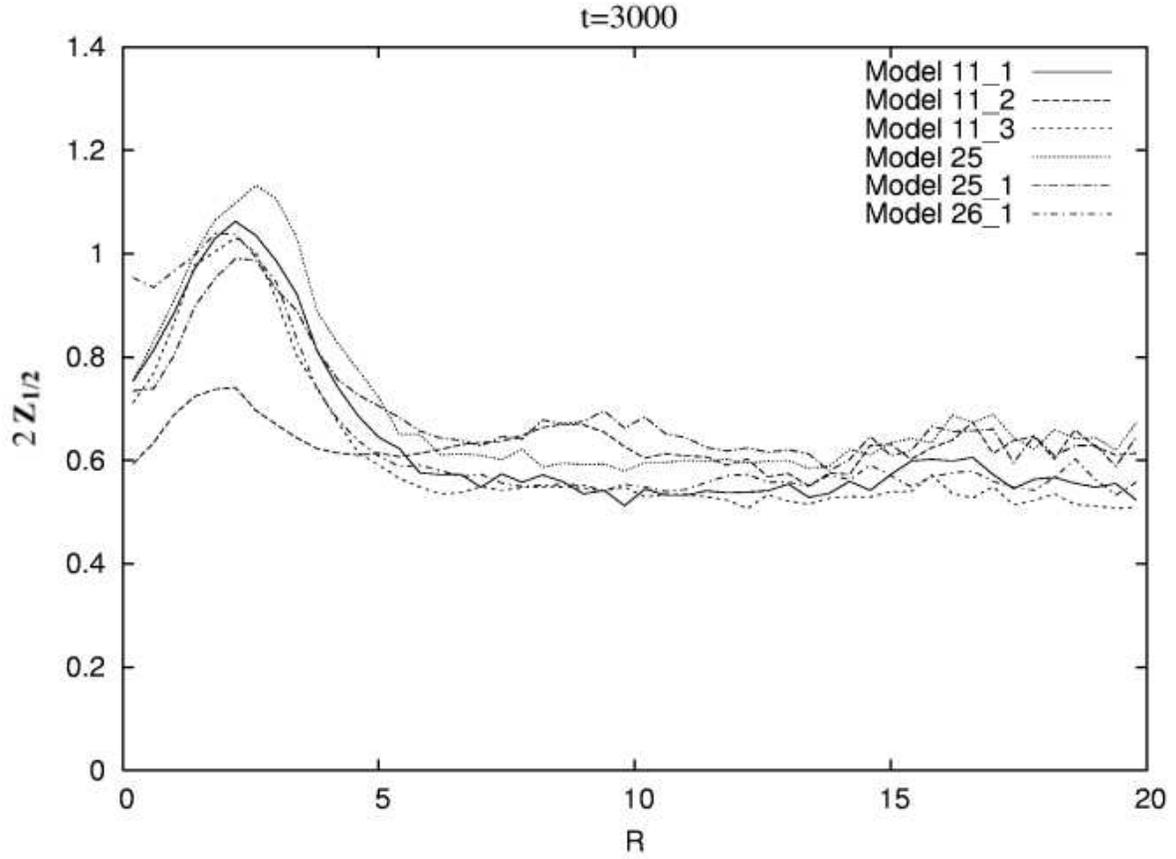,width=16cm}}
\caption[9]{Radial thickness profiles for the galaxy ($2z_{1/2}$) at
the time
$t = 3000$  for models that differ only by the initial thickness.
For all of the models,  $\mu = 0.6$ and $\mu_{\rm b} = 0$. Models~11\_1,
11\_2, and 11\_3 are different random realizations of a stellar system with
$z_0 = 0.1$; models~25 and 25\_1 are different random realizations of a
system with $z_0 = 0.2$; model~26\_1 is for $z_0 = 0.3$.}
\label{z0}
\end{figure}

The final disk thickness at fixed initial $Q_{\rm T}$ was determined only by
the relative mass of the spheroidal component and the contribution of the
bulge to this mass and did not depend on how far from stability the initial
state of the disk was chosen. The start from different initial disk
thicknesses led to models without any systematic differences between them.
Figure~\ref{z0} illustrates this result, which is similar to that obtained
in their numerical simulations by Sellwood and Merritt (1994).

Thus, our three-dimensional calculations are in good agreement with the
conclusion following from our analysis of the dispersion relation for a thin
disk that a bulge is an effective stabilizing factor during the growth of
bending instability. Moreover, since the initial bend is formed in the most
unstable central part of the galaxy (Sotnikova and Rodionov 2003), it is the
central regions that must be stabilized. This does not require a massive
dark halo; the presence of a compact spherical component like a bulge will
suffice. This suggests that the final thickness of the model galaxy depends
not only on the total mass of the spherical component, but also on the mass
distribution in it.

{\bf Barred galaxies.} The bending instability of bars is an effective
vertical disk heating mechanism for galaxies unstable against the formation
of a bar (Raha {\it et al.} 1991; Sotnikova and Rodionov 2003). In contrast
to the warp in the entire disk, the warp in the bar is formed not in the
central regions, but in the entire bar simultaneously (this is seen
particularly clearly in our color two-dimensional histograms of the warp
accessible at 
http://www.astro.spbu.ru/staff/seger/articles/warps\_2002/fig6\_web.html 
and 
http://www.astro.spbu.ru/staff/seger/articles/warps\_2002/fig7\_web.html). 
Therefore, the conclusion that a compact bulge during the growth of bending
instability in a bar will have the same effective stabilizing effect as
that for hot stellar disks is not obvious in advance.

In our simulations with $Q_{\rm T} = 1.5$ and $\mu \lsim 1.0$, the
warp in the bar was formed early, at $t \approx 800$. The presence of a
compact bulge eventually led to the formation of thinner disks, although the
effect itself was fairly complex.

Within $R < 1.5h$, the most prominent features of the bars at late
evolutionary stages were X-shaped structures. If the disk is viewed edge-on,
then they manifest themselves as a bulge with an appreciable extent in the
$z$ direction with boxy isophotes. In the region $R < 1.5h$ where the bar
dominated, we failed to reveal any distinct patterns in the model disk
thickness variations with increasing contribution of the bulge to the total
mass of the spheroidal component. In general, the presence of a bulge
``pushed forward''  the bar formation time and caused the saturation level
for the bending instability of a bar to lower. However, a further analysis
is required to completely understand the processes during the interaction
between a compact bulge and a bar.

\begin{figure}
\centerline{\psfig{file=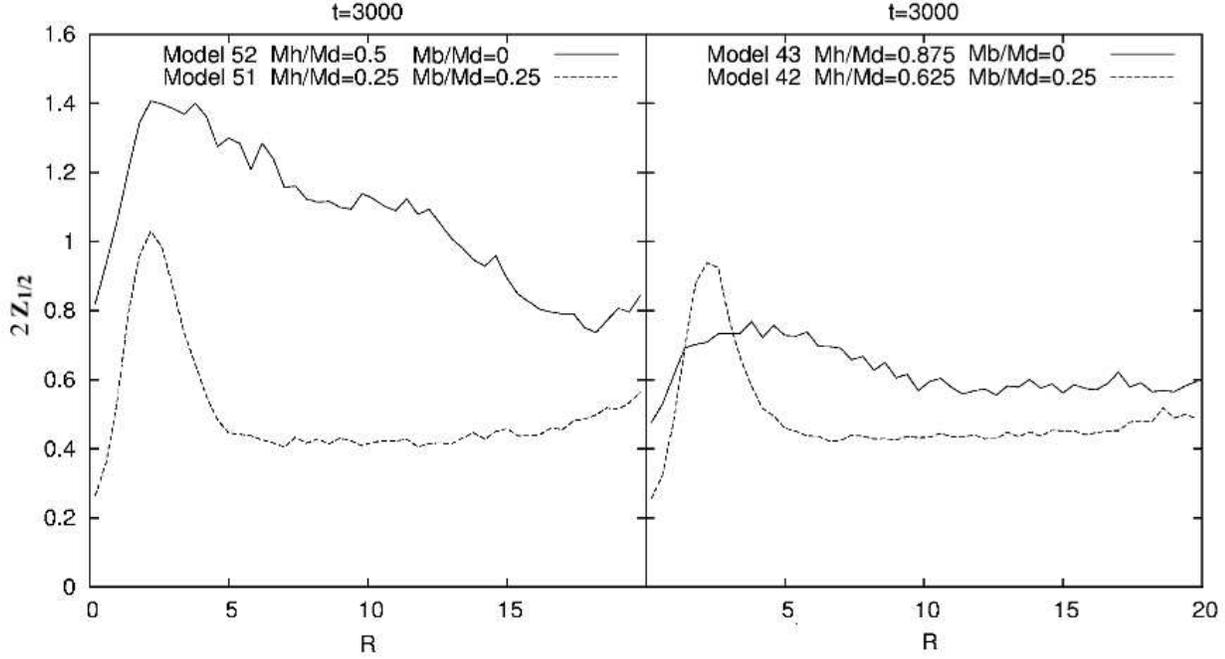,width=17cm}}
\caption[10]{Radial thickness profile for the galaxy ($2z_{1/2}$)
at the time
$t = 3000$ for models unstable against the growth of a bar
mode ($Q_{\rm T} = 1.5$): (a) models with $\mu = 0.5$ (model~52
with $\mu_{\rm b} = 0$, model~51 with $\mu_{\rm b} = 0.25$),
(b) models with $\mu = 0.875$ (model~43 with $\mu_{\rm b} = 0.25$, model~42
with $\mu_{\rm b} = 0.25$); the initial half-thickness
for all of the models is $z_0 = 0.1$.}
\label{thickness-b}
\end{figure}

As regards the peripheral regions of the disks ($R > 1.5h$), they were
always appreciably thinner in the models with a compact bulge at late
evolutionary stages. Figure~\ref{thickness-b} demonstrates this effect for
two groups of models with different bulge contributions:
models~52~and~51 with $\mu = 0.5$ and models~43~and~42 with $\mu = 0.875$.
The differences in thickness show up most clearly in the models with a small
relative mass of the spheroidal component. When $\mu$ increases, the disk
becomes very thin, as might be expected, and its thickness ceases to depend
on how the mass is distributed between the halo and the bulge.

\section{CONCLUSIONS}

A comparison of the conclusions that follow from a linear analysis with the
results of numerical simulations for three-dimensional disks shows that, in
contrast to homogeneous models, global bending modes with a wavelength
longer than the disk scale length can arise in inhomogeneous disks. If the
amplitude of the waves during the growth of instability increases
significantly, then they heat it up significantly in the vertical direction
as they pass through the entire disk. Hot disks are most unstable against
the growth of bending perturbations. An additional spheroidal component, for
example, a dark halo is a factor that stabilizes the bending perturbations.

Our additional qualitative analysis of the dispersion relation for
inhomogeneous models led us to new conclusions regarding the stabilization
conditions for the bending modes in stellar disks. These conclusions
were confirmed in our numerical simulations.

(1) Since the central regions of the disk (particularly if the disk is hot)
are most unstable, the conditions under which the growth of perturbations is
suppressed are determined not only by the mass of the spherical component,
but also by the density distribution in it. The suppressing effect is
enhanced with increasing concentration toward the center.

(2) The presence of a compact and moderately massive bulge in a galaxy
effectively prevents the growth of bending perturbations.

(3) It follows from an analysis of the entire set of our results that a more
accurate approach to estimating the dark halo mass from the observed relative
thickness of the stellar disk $z_0 / h$ in spiral galaxies is required.

\bc \bf ACKNOWLEDGMENTS
\ec

This work was supported by the Russian Foundation for Basic Research
(project no. 03-02-1752), the Federal ``Astronomy'' Program (project
no. 40.022.1.1.1101), and a grant from the President of Russia for support
of leading scientific schools (NSh-1088.2003.2).

\bc \bf REFERENCES
\ec

\hangindent=0.7cm \hangafter=1 \noindent
1. S. Araki, Ph. D. Thesis, Massachus. Inst. Tech.
(1985).

\hangindent=0.7cm \hangafter=1 \noindent
2. J. Barnes and P. Hut, Nature {\bf 324}, 446 (1986).

\hangindent=0.7cm \hangafter=1 \noindent
3. J. Binney and S. Tremaine, {\it Galactic Dynamics}
(Princeton Univ. Press, Princeton, 1987).

\hangindent=0.7cm \hangafter=1 \noindent
4. E. Griv, M. Gedalin, and Chi Yuan, Astrophys. J. {\bf 580},
L27 (2002).

\hangindent=0.7cm \hangafter=1 \noindent
5. E. Griv and Tzihong Chiueh, Astrophys. J. {\bf 503}, 186
(1998).

\hangindent=0.7cm \hangafter=1 \noindent
6. L. Hernquist, Astrophys. J., Suppl. Ser. {\bf 86}, 389
(1993).

\hangindent=0.7cm \hangafter=1 \noindent
7. P. C. van der Kruit and L. Searle, Astron. Astrophys.
{\bf 95}, 105 (1981).

\hangindent=0.7cm \hangafter=1 \noindent
8. R. M. Kulsrud, J.W.-K. Mark, and A. Caruso, Astrophys.
Space Sci. {\bf 14}, 52 (1971).

\hangindent=0.7cm \hangafter=1 \noindent
9. D. Merritt, Astron. J. {\bf 111}, 2462 (1996).

\hangindent=0.7cm \hangafter=1 \noindent
10. D. Merritt and J. A. Sellwood, Astrophys. J. {\bf 425}, 551
(1994).

\hangindent=0.7cm \hangafter=1 \noindent
11. E. A. Mikhailova, A. V. Khoperskov, and
S. S. Sharpak, {\it Stellar Dynamics --- from Clasic
to Modern} Ed. by L. P. Ossipkov and I. I. Nikiforov
(St. Petersburg State Univ. Press, St. Petersburg,
2001), p. 147.

\hangindent=0.7cm \hangafter=1 \noindent
12. P. A. Patsis, E. Athanassoula, P. Grosbol, and
Ch. Skokos Ch., Mon. Not. R. Astron.Soc. {\bf 335}, 1049
(2002).

\hangindent=0.7cm \hangafter=1 \noindent
13. V. L. Polyachenko and I. Sh. Shukhman, Pis'ma Astron.
Zh. {\bf 3}, 254 (1977) [Sov. Astron. Lett. {\bf 3}, 134
(1977)].

\hangindent=0.7cm \hangafter=1 \noindent
14. N. Raha, J. A. Sellwood, R. A. James, and F.D. Kahn,
Nature {\bf 352}, 411 (1991).

\hangindent=0.7cm \hangafter=1 \noindent
15. J. A. Sellwood, Astrophys. J. {\bf 473}, 733 (1996).

\hangindent=0.7cm \hangafter=1 \noindent
16. J. A. Sellwood and R. G. Carberg, Astrophys. J. {\bf 282},
61 (1984).

\hangindent=0.7cm \hangafter=1 \noindent
17. J. A. Sellwood and D. Merritt, Astrophys. J. {\bf 425}, 530
(1994).

\hangindent=0.7cm \hangafter=1 \noindent
18. N. Ya. Sotnikova and S. A. Rodionov, Pis'ma Astron. Zh. {\bf 29}, 
367 (2003) [Astron. Lett. {\bf 29}, 321 (2003)].

\hangindent=0.7cm \hangafter=1 \noindent
19. L. Spitzer, Astrophys. J. {\bf 95}, 325 (1942).

\hangindent=0.7cm \hangafter=1 \noindent
20. L. Spitzer and M. Schwarzshild, Astrophys. J. {\bf 114},
385 (1951).

\hangindent=0.7cm \hangafter=1 \noindent
21. L. Spitzer and M. Schwarzshild, Astrophys. J. {\bf 118},
106 (1953).

\hangindent=0.7cm \hangafter=1 \noindent
22. P. J. Teuben, ASP Conf. Ser. {\bf 77}, 398 (1995).

\hangindent=0.7cm \hangafter=1 \noindent
23. A. Toomre, Astrophys. J. 139, 1217 (1964).

\hangindent=0.7cm \hangafter=1 \noindent
24. A. Toomre, Geophys. Fluid Dyn., N66-46, 111
(1966).

\hangindent=0.7cm \hangafter=1 \noindent
25. Yao-Huan Tseng, Chinese Journ. Phys. {\bf 38}, 111
(2000).

\hangindent=0.7cm \hangafter=1 \noindent
26. H. Velasquez and S. D. M. White, Mon. Not. R.
Astron. Soc. {\bf 304}, 254 (1999).

\hangindent=0.7cm \hangafter=1 \noindent
27. I. W. Walker, J. Ch. Mihos, and L. Hernquist, Astrophys.
J. {\bf 460}, 121 (1999).

\hangindent=0.7cm \hangafter=1 \noindent
28. A. V. Zasov, D. V. Bizyaev, D. I. Makarov, and
N. V. Tyurina, Pis'ma Astron. Zh. {\bf 28}, 599 (2002) [Astron. Lett. {\bf
28}, 527 (2002)].

\hangindent=0.7cm \hangafter=1 \noindent
29. A. V. Zasov, D. I. Makarov, and E. A. Mikhailova,
Pis'ma Astron. Zh. {\bf 17}, 884 (1991) [Sov. Astron. Lett. {\bf 17}, 374
(1991)].

\begin{flushright} 
\it Translated by V. Astakhov
\end{flushright}

\newpage
 
\newpage
 
\newpage

\newpage

\newpage

\newpage

\newpage

\newpage

\newpage

\newpage

\end{document}